\documentclass[aps,pra,reprint,nofootinbib,twocolumn,superscriptaddress,showpacs,showkeys,longbibliography]{revtex4-1}

\usepackage{eurosym}
\usepackage{amsmath,amssymb,amstext}
\usepackage{ulem}
\usepackage[usenames,dvipsnames]{color}
\usepackage{graphicx}
\usepackage{braket}
\usepackage{natbib}
\usepackage{comment}
\usepackage{dcolumn}
\usepackage[english]{babel}
\usepackage{wasysym}
\usepackage{bm}
\usepackage[colorlinks,bookmarks=false,citecolor=blue,linkcolor=red,urlcolor=blue]{hyperref}

\usepackage{appendix}
\usepackage{CJK}
\begin{document}
\title{Static {non-linear Schr\"{o}dinger} equations for the achiral-chiral transitions  {of polar chiral molecules}}
\author{Chong Ye}
\affiliation{Beijing Computational Science Research Center, Beijing 100193, China}
\author{Quansheng Zhang}
\affiliation{Beijing Computational Science Research Center, Beijing 100193, China}
\author{Yong Li}\email{liyong@csrc.ac.cn}
\affiliation{Beijing Computational Science Research Center, Beijing 100193, China}
\affiliation{Synergetic Innovation Center for Quantum Effects and Applications, Hunan Normal University, Changsha 410081, China}

\begin{abstract}
  In the mean-field theory, the stabilization of  {polar} chiral molecules is understood as a quantum phase transition where the mean-field ground state of molecules changes from the achiral eigenstate of the molecular Hamiltonian to one of the degenerated chiral states as the increase of the intermolecular interaction. Starting from the many-body Hamiltonian of the molecular gases with electric dipole-dipole interactions, we give the \textit{static} {non-linear Schr\"{o}dinger} equations without free parameters to explore the achiral-chiral transitions of  {polar} chiral molecules.
  We find that the  {polar} chiral molecules of different species can be classified into two categories:
  {At the critical point for the achiral-chiral transition, the mean-field ground state changes continuously in one category, and changes discontinuously in the other category. We further give the mean-field phase diagram of the achiral-chiral transitions for both two categories.}

\end{abstract}
%
\date{\today}
\maketitle
\section{Introduction}
The stabilization of chiral molecules
is an old but important problem in molecular quantum mechanics.
According to quantum theory, the molecules might be
expected to stay in the ground state $|+\rangle$ of the parity-invariant molecular Hamiltonian in the space of the  {chirality (related vibrational)} degree of freedom. However, chiral molecules commonly stay stably in the localized left-handed state $|L\rangle$ or right-handed state $|R\rangle$, which are respectively the symmetric and anti-symmetric superpositions of the ground state $|+\rangle$ and the first excited state $|-\rangle$ of the parity-invariant molecular Hamiltonian. The left- and right-handed states are chiral in the sense that $|L\rangle=\mathcal{\hat{T}}|R\rangle$ with the parity operator $\mathcal{\hat{T}}$. This stabilization problem of chiral molecules has been explained previously with the exceedingly long tunneling time~\cite{Z.Phys.43.805} between $|L\rangle$ and $|R\rangle$ and/or introducing party-violating terms~\cite{PV1,PV2,PV3,PV4,PV5,PV6} in the molecular
Hamiltonian, by considering the single molecule as an isolated system. Later on, it is found that the above mechanisms are not sufficient to explain the observed stabilization of some kinds of chiral molecules~\cite{PR.75.1450,PR.76.1423,PRL.40.980,MC1,MC2,JCP.112.8743,PRL.103.023202,PRA.88.032504,PCCP.13.17130}.

In fact, isolated molecules do not exist in nature. The effects of the environment must be taken into consideration in realistic physical systems. The intermolecular interaction is one of the effects of the environment attributing to the stabilization of chiral molecules~\cite{PR.75.1450,PR.76.1423,PRL.40.980,MC1,MC2}. Many approaches have been proposed to quantitatively deal with the effect of the intermolecular interaction, where the most well-known ones are the mean-field theory~\cite{JCP.112.8743,PRA.91.022709,PRL.88.123001} and the decoherence theory~\cite{PRL.103.023202,PRA.84.062115,PCCP.14.9214,PRA.93.063612,PRA.88.032504,PRA.86.042111,PRA.81.052701}.  There are also proposals combining both the mean-field and the decoherence theories to study the stabilization of chiral molecules~\cite{PCCP.13.17130,CPL.516.29}. According to the mean-field theory, the stabilization of chiral molecules is the result of a quantum phase transition from an achiral phase to a chiral phase~\cite{PRA.91.022709,PRL.88.123001}, namely the achiral-chiral transition. According to the decoherence theory, the stabilization of chiral molecules can be understood~\cite{PRL.103.023202} in analogy to the quantum Zeno effect~\cite{JMP.18.756} when the environment behaves as continuously monitoring the molecular state.

Using the mean-field theory, Jona-Lasinio and coworkers~\cite{PRA.91.022709,PRL.88.123001}
introduced an ingenious model in the  {chirality degree of freedom} to explore the static and {time-dependent} problems of  {polar} chiral molecules of $\mathcal{C}_2$ symmetry, which have $zero$ permanent electric dipole momentum in an achiral state  {and $nonzero$ permanent electric dipole momentum in a chiral state.}
{They quantitatively describes, without free parameters, the effect of the intermolecular interaction in the stabilization of chiral molecules by giving simply the effective interaction between $i$-th and $j$-th molecules as the form $\sigma^{z}_{i}\sigma^{z}_{j}$ with $\sigma^{z}$ given in the basis $\{|L\rangle,|R\rangle\}$. However, this form of the effective interaction {is not sufficient} to describe {polar} chiral molecules of $\mathcal{C}_{1}$ symmetry~\cite{PCCP.13.17130},  {which have  $nonzero$ permanent electric dipole momentums in both an achiral state and a chiral state. Thus, such chiral molecules have $nonzero$ interaction energy in an achiral state according to the Keesom's theory~\cite{Phys.Z.22.129}.}


{In this manuscript, we re-investigate the stabilization of the  {polar} chiral molecules by means of mean-field theory directly from the standard form of the electric dipole-dipole interactions. Since the dipole-dipole interactions are directly related to the positions and orientations of the molecules, the many-body state of the system in the spatial (position) and rotational (orientation) degrees of freedom is considered thoroughly via assuming that the two-particle density follows the Bolztmann distribution~\cite{Book1} with respect to the electric dipole-dipole interaction.
Summing over all possible molecular distances and all possible dipole orientations with the help of the Bolztmann distribution and using the standard method of Lagrange
multipliers, we arrive our static non-linear Schr\"{o}dinger equations
in the space of the  {chirality degree of freedom}. }

{Self-consistently solving the non-linear eigenvalue problem,} we find that the  {polar} chiral molecules can be classified into two categories. In one category, the mean-field ground state changes continuously from the achiral ground state of the parity-invariant molecular Hamiltonian to one of two degenerated chiral states with the increase of the intermolecular interaction. In the other category, the mean-field ground state changes discontinuously {from the achiral ground state to one of the two degenerated chiral ones}.
{In contrast, only the continuous transitions have been predicted in previous works~\cite{
PRA.91.022709,PRL.88.123001}.}

\section{Model}
\subsection{Many-body Hamiltonian}
Without taken the kinetic and rotational energies of single molecules into
consideration~\cite{JCP.112.8743,PRL.88.123001,
PRA.91.022709,PRL.103.023202,PRA.84.062115,PCCP.14.9214,PRA.93.063612,
PCCP.13.17130}, the many-body Hamiltonian for the system of $N$ molecules is
\begin{align}\label{MBH}
\hat{H}=\sum^{N}_{i=1}(-\frac{\omega}{2}|L\rangle_{ii}\langle R|+h.c.)
+\sum^{N}_{i=1}\sum^{N}_{j=i+1}\hat{V}_{ij},
\end{align}
where the first term is the parity-invariant
molecular Hamiltonian given in the basis $\{|L\rangle,|R\rangle\}$ and
the second term describes the standard electric dipole-dipole interaction between the $i$-th and $j$-th molecules located at $\bm{r}^{s}_{i}$ and $\bm{r}^{s}_{j}$ with the form
\begin{align}\label{DDI}
&\hat{V}_{ij}
=\frac{\hat{\bm{\mu}}^{s,i}\cdot\hat{\bm{\mu}}^{s,j}-3(\hat{\bm{\mu}}^{s,i}\cdot \bm{r}^{s}_{ij})(\hat{\bm{\mu}}^{s,j}\cdot \bm{r}^{s}_{ij})r^{-2}_{ij}} {4\pi\varepsilon_0 r^3_{ij}}.
\end{align}
Here $\hat{\bm{\mu}}^{s,i}$ is the electric dipole operator of the $i$-th molecule in the space-fixed frame, the notation ``$s$'' indicates the space-fixed
frame, and $r_{ij}=|\bm{r}^{s}_{ij}|$. $\bm{r}^{s}_{ij}$ ($\equiv \bm{r}^{s}_{i}-\bm{r}^{s}_{j}$ can be expressed as
\begin{align}\label{rsij}
\bm{r}^{s}_{ij}=r_{ij}\sum_{\sigma=0,\pm1}\sqrt{\frac{4\pi}{3}}
Y_{1\sigma}(\tilde{\Omega}_{\bm{r}_{ij}^{s}})\bm{e}^{s}_{\sigma},
\end{align}
where $\tilde{\Omega}_{\bm{r}^{s}}=(\alpha_{\bm{r}^{s}},\beta_{\bm{r}^{s}})$
are the solid angles of $\bm{r}^{s}$ and $Y_{1\sigma}$ are the spherical harmonics. Here
$\bm{e}^{s}_{0}=\bm{e}_Z$ and $\bm{e}^{s}_{\pm1}=(\mp\bm{e}_X-i\bm{e}_Y)/\sqrt{2}$
with the coordinations in the space-fixed frame $(X,Y,Z)$.

The components of the electric dipole operator $\hat{\bm{\mu}}^{s}$ {of a general chiral molecule} in the space-fixed frame can be obtained by a rotation
from the molecular frame~\cite{JCP.137.044313} via
\begin{align}\label{MU1}
\hat{\mu}_{\sigma}^{s}
=\sum_{\sigma^{\prime}=0,\pm1}
D^{1\ast}_{\sigma\sigma^{\prime}}
(\alpha,\beta,\gamma)
\hat{\mu}^{m}_{\sigma^{\prime}}
\end{align}
with $\hat{\mu}^{s}_{0}=\hat{\mu}^{s}_Z$ and $\hat{\mu}^{s}_{\pm1}=
(\mp{\hat{\mu}}^{s}_X - i{\hat{\mu}}^{s}_Y)/\sqrt{2}$.
The index ``$m$'' indicates the molecular frame and ``$\ast$'' denotes taking conjugate complex. $D^{1}$ is the rotation matrix in three dimensions. Here $\Omega=(\alpha,\beta,\gamma)$ are the Euler angles denoting the orientation of the molecule. 
$\hat{\mu}^{m}_{\sigma^{\prime}}$ are the components of the electric dipole $\hat{\bm{\mu}}^{m}$ in the molecular frame with $\hat{\mu}^{m}_{0}=\hat{\mu}^{m}_z$ and $\hat{\mu}^{m}_{\pm1}=
(\mp{\hat{\mu}}^{m}_x - i{\hat{\mu}}^{m}_y)/\sqrt{2}$.
Here $x$, $y$, and $z$ are, respectively, the principal (inertial) axes of the molecule in the molecular frame.

\subsection{Static {non-linear Schr\"{o}dinger} equations}
We assume the $N$-molecule ($N\gg1$) system is described by the density matrix{
\begin{align}
\rho&=\rho_{sr}(\bm{r}^{s}_{1},...,\bm{r}^{s}_{N};
\Omega_{1},...,\Omega_{N}; \lambda_{1},...,\lambda_{N}) \otimes \prod^{N}_{i=1}|\lambda_{i}\rangle\langle\lambda_{i}|,
\end{align}
where $\rho_{sr}$ describes the density matrix of the system in the spatial and rotational degrees of freedom with $\bm{r}^{s}_{i}$ and $\Omega_{i}$ denoting the position and orientation of the $i$-th molecule.
 {Indeed, a molecule in general has many vibrational degrees of freedom. We only refer to the one relating to the two chiral states (called as chirality degree of freedom) and assume that all other vibrational degrees of freedom are frozen.}
{Since the electric dipole-dipole interactions couple the  {chirality}, spatial, and rotational degrees of freedom, it is natural that $\rho_{sr}$ is dependent on the states of molecules in the chirality degree of freedom $|\lambda_{i}\rangle$.}} {With the framework of mean-field theory, we replace the state {in the chirality degree of freedom} $|\lambda_{i}\rangle$ with a mean-field state of }
\begin{align}\label{State}
|\lambda\rangle=\varphi_{\lambda,L}|L\rangle
+\varphi_{\lambda,R}|R\rangle.
\end{align}{
Correspondingly, we will use $\rho_{sr}(\bm{r}^{s}_{1},...,\bm{r}^{s}_{N};
\Omega_{1},...,\Omega_{N}; \lambda)$ to replace $\rho_{sr}(\bm{r}^{s}_{1},...,\bm{r}^{s}_{N};
\Omega_{1},...,\Omega_{N}; \lambda_{1},...,\lambda_{N})$.}

Then, the energy of the system, $E\equiv\mathrm{Tr}(\rho\hat{H})$, can be expressed as
\begin{align}
E\simeq-N\frac{\omega}{2}(\varphi^{\ast}_{\lambda,L}\varphi_{\lambda,R}+c.c.)
+\frac{N(N-1)}{2}g(\lambda),
\end{align}
where the {mean-field} two-particle interaction energy is{
\begin{align}
g(\lambda)\equiv\mathrm{Tr}(\rho_{sr}\hat{V}_{12})=\mathrm{Tr}(\rho_{sr}\langle \lambda,\lambda|\hat{V}_{12}
| \lambda,\lambda\rangle).
\end{align}
Here $|\lambda,\lambda\rangle=|\lambda_{1}\rangle\otimes|\lambda_{2}\rangle$ with $|\lambda_{1,2}\rangle=|\lambda\rangle$.} And $V_{12}(\lambda)\equiv\langle \lambda,\lambda|\hat{V}_{12}
| \lambda,\lambda\rangle$
is an operator in the space of the spatial and rotational degrees of freedom.
Explicitly, we have
\begin{align}\label{DDI2}
&V_{12}(\lambda)
=-\frac{1}{4\pi\varepsilon_0 r^3_{12}}
\sum_{\sigma^{\prime}_{1},\sigma^{\prime}_{2}=0,\pm1}
{\mu}^{m}_{\lambda,\sigma^{\prime}_{1}}(\hat{\mu}^{m}_{\lambda,\sigma^{\prime}_{2}})^{\ast}\times\nonumber\\
&[\sum_{\sigma_{1},\sigma_{2}=0,\pm1}4\pi
D^{1}_{\sigma_{1}\sigma^{\prime}_{2}}(\Omega_{2})
D^{1\ast}_{\sigma_{1}\sigma^{\prime}_{1}}(\Omega_{1})
Y_{1\sigma_{1}}(\tilde{\Omega}_{\bm{r}^{s}_{12}})
Y^{\ast}_{1\sigma_{2}}(\tilde{\Omega}_{\bm{r}^{s}_{12}})\nonumber\\
&-\sum_{\sigma=0,\pm1}
D^{1\ast}_{\sigma\sigma^{\prime}_{1}}(\Omega_{1})
D^{1}_{\sigma\sigma^{\prime}_{2}}(\Omega_{2})]
\end{align}
with $
\mu^{m}_{\lambda,\sigma^{\prime}}\equiv\langle\lambda|\hat{\mu}^{m}_{\sigma^{\prime}}|\lambda\rangle$.

Further the mean-field two-particle interaction energy approximates to
\begin{align}\label{G1}
g(\lambda)=\mathrm{Tr}_{\{1,2\}}[V_{12}(
\lambda)\rho^{sr}_{12}(\bm{r}_1,\bm{r}_2;\Omega_1,\Omega_2;\lambda)]
\end{align}
with the two-particle density matrix
\begin{align}
\rho^{sr}_{12}(\bm{r}_1,\bm{r}_2;\Omega_1,\Omega_2;\lambda)\equiv \mathrm{Tr}_{\{3,..,N\}}[\rho_{sr}],
\end{align}
where $\mathrm{Tr}_{\{1,2\}}$ means an integral over $\bm{r}^{s}_{1}$, $\bm{r}^{s}_{2}$, $\Omega_1$, and $\Omega_{2}$, and $\mathrm{Tr}_{\{3,..,N\}}$ means an integral over all coordinates and Euler angles except $\bm{r}^{s}_{1}$, $\bm{r}^{s}_{2}$, $\Omega_1$, and $\Omega_{2}$.

The two-particle density matrix can be approximately written as~\cite{Book1}
\begin{align}\label{rho2}
&\rho^{sr}_{12}(\bm{r}_1,\bm{r}_2;\Omega_1,\Omega_2)
\simeq\frac{1}{\mathcal{Z}}
\exp[-\frac{V_{12}(\lambda)}
{k_B T}]
\end{align}
with the Boltzmann constant $k_{B}$ and the normalization constant $\mathcal{Z}$.
Here we have assumed the Boltzmann distribution of the two-particle density in the spatial and rotational degrees of freedom~\cite{Book1}.
Since $V_{12}$ and $\rho^{sr}_{12}$ are functions of $\bm{r}^{s}_{12}$, we make variable substitution
as $\int d^3\bm{r}^{s}_{1} d^3\bm{r}^{s}_{2}...=\int d^3\bm{r}^{s}_{1} d^3\bm{r}^{s}_{12}...$ and integrate over $\bm{r}^{s}_{1}$ first. Then we can get
\begin{align}\label{gintegrate}
g(\lambda)=\frac{1}{\mathcal{V}}\int d^3\bm{r}^{s}_{12}d\Omega_{1}d\Omega_{2}V_{12}\rho^{sr}_{12}
\end{align}
with the volume of the gas $\mathcal{V}$.
Assuming $V_{12}/(k_B T)\ll1$ and applying the Taylor expansion to $g(\lambda)$,
we have
\begin{align}\label{O2}
g(\lambda)
&=-\frac{512\pi^5}{3}\frac{1}{\mathcal{Z}}\frac{\mathcal{V}}{k_{B}T}\int_{r_{12}>d} r^2_{12}dr_{12}\frac{|\bm{\mu}^{m}_{\lambda}|^4}{(4\pi\varepsilon_{0}r^3_{12})^2}
\nonumber\\
&=-\frac{1}{N}\frac{ P|\bm{\mu}^{m}_{\lambda}|^4}
{18\pi(\varepsilon_0k_BT)^2d^3}
\end{align}
with $P$ the pressure of the gas and $d$ the molecular collision diameter. We have used $\int d^3\bm{r}_{1} d^3\bm{r}_{2}d\Omega_{1}d\Omega_{2}V_{12}=0$,
$\mathcal{Z}\simeq 64\pi^4\mathcal{V}^2$,
and the approximation
of ideal gas $P\mathcal{V}\simeq Nk_BT$. We note that ${|\bm{\mu}^{m}_{\lambda}|^4}/{(4\pi\varepsilon_{0}r^3_{12})^2}$
is the Van der Waals potential between
two electric dipoles, namely the Keesom
interaction~\cite{Phys.Z.22.129}.

The average energy of a molecule in the system, $\varepsilon(\lambda)\equiv E/N$,
is approximately given as
\begin{align}\label{vpsl}
\varepsilon(\lambda)
=-\frac{\omega}{2}(\varphi^{\ast}_{\lambda,L}\varphi_{\lambda,R}+c.c.)
-\frac{P|\bm{\mu}^{m}_{\lambda}|^4}{36\pi(\varepsilon_0 k_{B}T)^2d^3}
\end{align}
by neglecting the terms of order $1/N$ in the large $N$ limit.

For a pair of left-handed and right-handed chiral states $|L\rangle$ and $|R\rangle$, we have
$\bm{\mu}^{m}_{L}\equiv\langle L|\bm{\hat{\mu}}^{m}|L\rangle$ and
$\bm{\mu}^{m}_{R}\equiv\langle R|\bm{\hat{\mu}}^{m}|R\rangle$.
It is well known that their components satisfy~\cite{XX1,XX2,XX10,XX11,XX21,XX3,XX4,XX5}
\begin{align}
\mu^{m}_{L,x}\mu^{m}_{L,y}\mu^{m}_{L,z}=-\mu^{m}_{R,x}\mu^{m}_{R,y}\mu^{m}_{R,z}.
\end{align}

\subsubsection{ {Polar chiral molecules of $\mathcal{C}_{1}$ symmetry}}
For  {polar} chiral molecules of $\mathcal{C}_{1}$ symmetry {whose three components of the electric-dipole along the three principle axes are $nonzero$}, we can assume that the components obey~\cite{PRE.59.2105,JCP.113.7}
\begin{align}\label{MU2}
&\mu^{m}_{L,x}=\mu^{m}_{R,x}=\mu^{m}_{x}\ne 0,\nonumber\\
&\mu^{m}_{L,y}=\mu^{m}_{R,y}=\mu^{m}_{y}\ne 0,\nonumber\\
&\mu^{m}_{L,z}=-\mu^{m}_{R,z}=\mu^{m}_{z}\ne 0.
\end{align}
With Eq.~(\ref{MU2}), we have
\begin{align}\label{MU3}
&\mu^{m}_{\lambda,\pm1}\equiv\langle\lambda|\hat{\mu}^{m}_{\pm1}|\lambda\rangle=\frac{\mp\mu^{m}_{x}- i\mu^{m}_{y}}{\sqrt{2}},\nonumber\\
&\mu^{m}_{\lambda,0}\equiv\langle\lambda|\hat{\mu}^{m}_{0}|\lambda\rangle
=\mu^{m}_{z}(|\varphi_{\lambda,L}|^2-|\varphi_{\lambda,R}|^2).
\end{align}

By means of Eqs.~(\ref{vpsl},\ref{MU3}) and the condition $|\varphi_{\lambda,L}|^2+|\varphi_{\lambda,R}|^2=1$, we can get the static {non-linear Schr\"{o}dinger} equations for the system in the space of the {chirality degree of freedom} via the method of Lagrange multipliers as
\begin{align}\label{EoS}
&-\frac{\omega}{2}\varphi_{\lambda,L}+US^3_z(\lambda)\varphi_{\lambda,R}
+GS_z(\lambda)\varphi_{\lambda,R}=\eta\varphi_{\lambda,R},\nonumber\\
&-\frac{\omega}{2}\varphi_{\lambda,R}-US^3_z(\lambda)\varphi_{\lambda,L}
-GS_z(\lambda)\varphi_{\lambda,L}=\eta\varphi_{\lambda,L},
\end{align}
where the eigenvalue $\eta$ is the chemical potential, and{
\begin{align}
&S_z(\lambda)\equiv|\varphi_{\lambda,L}|^2-|\varphi_{\lambda,R}|^2,\\
&U=\frac{(\mu^{m}_{z})^4P}
{36\pi(\varepsilon_{0}k_{B}T)^2d^3},\\
&G=\frac{(\mu^{m}_{z}\mu^{m}_{\perp})^2P}
{36\pi(\varepsilon_{0}k_{B}T)^2d^3}
\end{align}}
with $\mu^{m}_{\perp}\equiv\sqrt{(\mu^{m}_{x})^2+(\mu^{m}_{y})^2}$.

{For  {polar} chiral molecules of $\mathcal{C}_{1}$ symmetry in an achiral state ($\varphi_{\lambda,L}=\pm\varphi_{\lambda,R}$),
they have $nonzero$ components of the permanent electric dipole momentum, $\mu^{m}_{\lambda,\pm1}\ne 0$. This means the two-particle interaction energy is $nonzero$. In contrast, one would obtain $zero$ two-particle interaction energy for  {polar} chiral molecules of $\mathcal{C}_{1}$ symmetry in an achiral state according to the model in Refs.~\cite{PRA.91.022709,PRL.88.123001}. This contradicts the Keesom's theory~\cite{Phys.Z.22.129}, where the two-particle interaction energy for  {polar} chiral molecules of $\mathcal{C}_{1}$ symmetry in an achiral state is $nonzero$. Thus, the model in Refs.~\cite{PRA.91.022709,PRL.88.123001} is not available for  {polar} chiral molecules of $\mathcal{C}_{1}$ symmetry, as also pointed out in Ref.~\cite{PCCP.13.17130}.}

\subsubsection{ {Polar chiral molecules of $\mathcal{C}_2$ symmetry}}
For polar chiral molecules of $\mathcal{C}_2$ symmetry, {only one of the three components of the electric-dipole along the three principle axes is $nonzero$. In the problem of the stabilization of {polar} chiral molecules, the  {polar} chiral molecules of $\mathcal{C}_2$ symmetry can be considered as a special case of  {polar} chiral molecules of $\mathcal{C}_2$ symmetry.} We can assume that the components obey~\cite{Ha}
\begin{align}
&\mu^{m}_{L,x}=\mu^{m}_{R,x}=\mu^{m}_{x}= 0,\nonumber\\
&\mu^{m}_{L,y}=\mu^{m}_{R,y}=\mu^{m}_{y}= 0,\nonumber\\
&\mu^{m}_{L,z}=-\mu^{m}_{R,z}=\mu^{m}_{z}\ne 0.
\end{align}
Then we have
\begin{align}
&\mu^{m}_{\lambda,\pm1}\equiv\langle\lambda|\hat{\mu}^{m}_{\pm1}|\lambda\rangle=0\nonumber\\
&\mu^{m}_{\lambda,0}\equiv\langle\lambda|\hat{\mu}^{m}_{0}|\lambda\rangle
=\mu^{m}_{z}(|\varphi_{\lambda,L}|^2-|\varphi_{\lambda,R}|^2),
\end{align}
and get the static {non-linear Schr\"{o}dinger} equations in the space of the  {chirality degree of freedom} as
\begin{align}\label{EoS11}
&-\frac{\omega}{2}\varphi_{\lambda,L}+US^3_z(\lambda)\varphi_{\lambda,R}
=\eta\varphi_{\lambda,R},\nonumber\\
&-\frac{\omega}{2}\varphi_{\lambda,R}-US^3_z(\lambda)\varphi_{\lambda,L}
=\eta\varphi_{\lambda,L}.
\end{align}
In fact, Eq.~(\ref{EoS11}) is the special case of Eq.~(\ref{EoS}) when $G=0$.

{In Refs.~\cite{PRA.91.022709,PRL.88.123001}, the non-linear terms are proportional to $S_z(\lambda)$ since they have used the effective interaction of the form $\sigma^{z}_{i}\sigma^{z}_{j}$ to describe the electric dipole-dipole interactions. However, we obtain a different form of non-linear Sch\"{o}dinger equations~(\ref{EoS11}) with the non-linear terms proportional to $S^{3}_z(\lambda)$ by using the standard form of the electric dipole-dipole interactions.}

{We {would like to remark }that the very central point of the manuscript is the appearance of the term $|\bm{\mu}^{m}_{\lambda}|^4$ in Eq.~(\ref{O2}), where $|\bm{\mu}^{m}_{\lambda}|^{2}$ will be introduced twice in the mean-field two-particle interaction energy. For the first time, it is introduced in $V_{12}(\lambda)$ of Eq.~(\ref{DDI2}), which gives the mean-field two-particle interaction energy of two orientated molecules of Euler angler $\Omega_1$ and $\Omega_2$. For the second time, it is
introduced in making the sum over all possible $\Omega_1$ and $\Omega_2$ in Eq.~(\ref{gintegrate}) via assuming the Boltzmann distribution of the two-particle density in Eq.~(\ref{rho2}).
{It is the appearance of $|\bm{\mu}^{m}_{\lambda}|^4$ in Eq.~(\ref{O2}) that induces the non-linear terms proportional to $S^{3}_z(\lambda)$ in our non-linear Sch\"{o}dinger equations~(\ref{EoS}) and (\ref{EoS11}).} }

\section{Achiral-chiral transition}
Starting from the many-body Hamiltonian~(\ref{MBH}), we have obtained the static {non-linear Schr\"{o}dinger} equations for all the polar chiral molecules with inversion symmetry in the space of the {chirality degree of freedom}. In the following, we will explore the achiral-chiral transitions via solving the nonlinear eigenvalue problem associated with Eq.~(\ref{EoS}). The coefficients $\varphi_{\lambda,L}$ and $\varphi_{\lambda,R}$ can be chosen real.
With Eq.~(\ref{EoS}), we have
\begin{align}\label{EoS1}
S_z(\lambda)\{4\varphi_{\lambda,L}\varphi_{\lambda,R}
[US^2_z(\lambda)+G]-\omega \}=0.
\end{align}
Once the solutions are found, the corresponding eigenvalues (chemical potential) are given by
\begin{align}\label{EoE1}
\eta=-\omega\varphi_{\lambda,L}\varphi_{\lambda,R}
-US^4_z(\lambda)-GS^2_z(\lambda).
\end{align}

\begin{figure}[h]
  \centering
  \includegraphics[width=1\columnwidth]{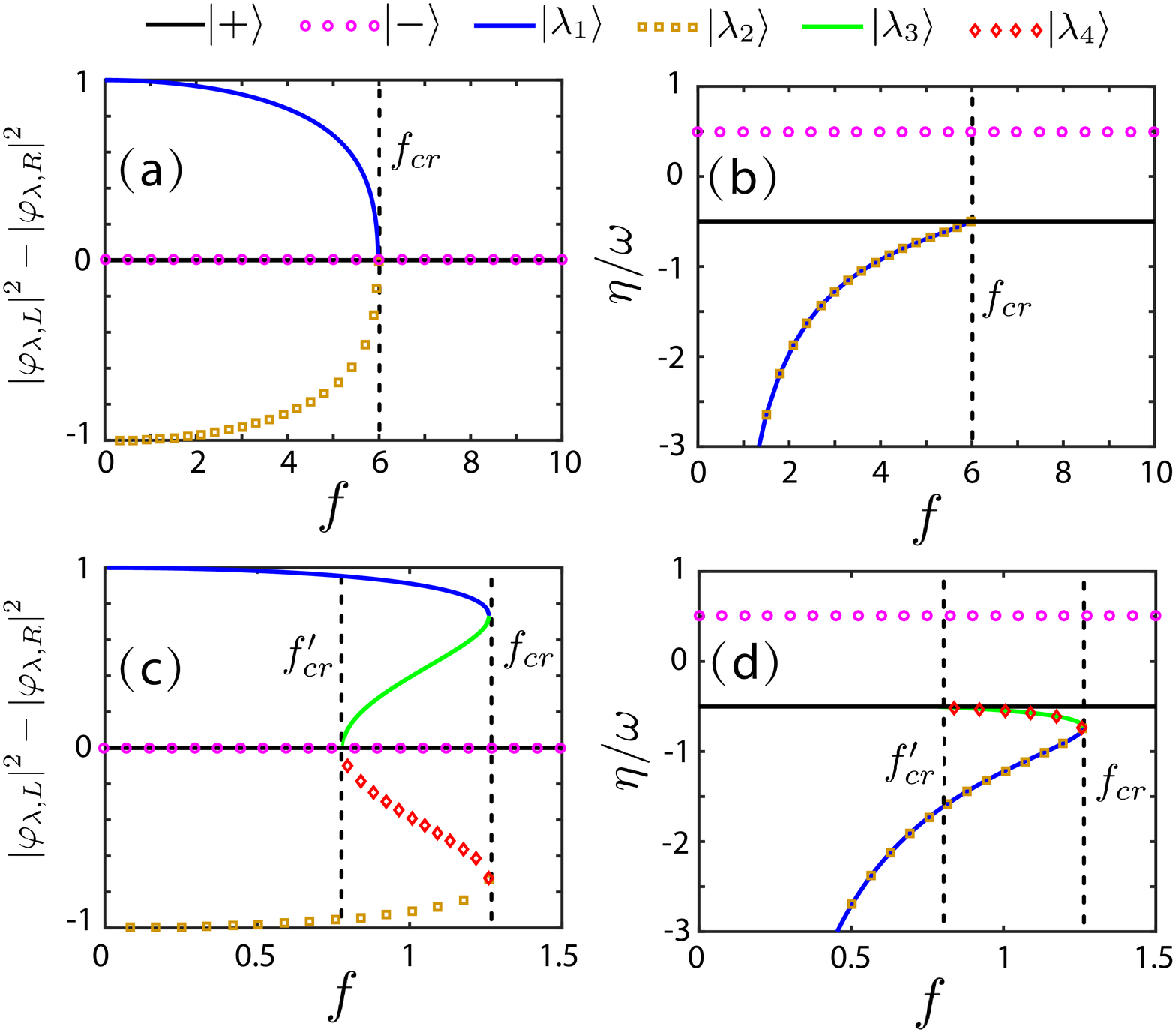}\\
  \caption{Achiral-chiral transitions.
  (a) and (b) present the achiral-chiral transition in category (I) with
  $\mathcal{G}=3$. (c) and (d) present the achiral-chiral transition in category (II) with $\mathcal{G}=0.4$. Here $|\pm\rangle$ are the two achiral eigen-states. The other possible eigen-states $|\lambda_1\rangle$, $|\lambda_2\rangle$, $|\lambda_3\rangle$ and $|\lambda_4\rangle$ are chiral in the sense $|\lambda_1\rangle=\hat{\mathcal{T}}|\lambda_2\rangle$ and $|\lambda_3\rangle=\hat{\mathcal{T}}|\lambda_4\rangle$ with the parity operator $\hat{\mathcal{T}}$. Here, we have $f\equiv\omega/U$ and $\mathcal{G}\equiv G/U$.  }\label{Fig3}
\end{figure}
There are always two solutions to Eq.~(\ref{EoS1}) when $S_{z}(\lambda)=0$. Up to an irrelevant sign, these two solutions correspond to the achiral eigenstates
\begin{align}
|\pm\rangle=\frac{1}{\sqrt{2}}(|L\rangle\pm|R\rangle)
\end{align}
with the corresponding eigenvalues $\eta_{\pm}=\mp\omega/2$. They are, respectively, the ground and first excited eigenstates of the parity-invariant molecular Hamiltonian.

{We note that
$\mathcal{G}\equiv G/U=(\mu^{m}_{\perp}/\mu^{m}_{z})^2$ is determined by the species of polar chiral molecules
and thus it can not be given arbitrarily. For the  {polar} chiral molecules of $\mathcal{C}_{2}$ symmetry, $\mathcal{G}=0$. For the  {polar} chiral molecules of $\mathcal{C}_{1}$ symmetry,
$\mathcal{G}\ne0$. For example, $\mathcal{G}\simeq2.7$ for propylene oxide~\cite{JPC.83.1457} and $\mathcal{G}\simeq0.46$ for solketal~\cite{XX2}.
For convenience of discussion in the following, we will use $\mathcal{G}$ as a tunable parameter.}

For a given value of $\mathcal{G}$, there will be further chiral solutions to Eq.~(\ref{EoS1}) when $f$ ($\equiv\omega/U$) is smaller
than a critical value $f_{cr}$ as shown in Fig.~\ref{Fig3}.
The corresponding chiral states have lower energies than the achiral state $|+\rangle$.
The decrease of $f$ will give rise to the achiral-chiral transition.
Changing $\mathcal{G}$, we find that the achiral-chiral transitions
can be divided into two categories: In category (I), the mean-field ground state changes continuously from the achiral state $|+\rangle$ to a chiral state with the decrease of $f$; in category (II), the mean-field ground state changes discontinuously at the critical point of $f_{cr}$.

In Fig.~\ref{Fig3}~(a) and Fig.~\ref{Fig3}~(b), we choose $\mathcal{G}=3$ to show
the typical behaviors of the achiral-chiral transitions in
category (I). When $f$
decreases from the region $f>f_{cr}$ to the region $f<f_{cr}$,
the mean-field ground state will change continually from
the achiral state $|+\rangle$ to one of the two degenerated
chiral states $|\lambda_{1}\rangle$ and
$|\lambda_{2}\rangle$.
They are chiral in the sense~\cite{PRA.91.022709} that
$|\lambda_{1}\rangle=\hat{\mathcal{T}}|\lambda_{2}\rangle$ with
the parity operator $\hat{\mathcal{T}}$.
In the limit $f\ll 1$ ($\omega\ll U$), we find that
$|\lambda_{1}\rangle$ and $|\lambda_{2}\rangle$
approach the localized states $|L\rangle$
and $|R\rangle$, respectively.
In category (II), the typical behaviors are shown
in Fig.~\ref{Fig3}~(c) and Fig.~\ref{Fig3}~(d)
with $\mathcal{G}=0.4$. When $f$
decreases from the region $f>f_{cr}$ to the region $f<f_{cr}$,
the mean-field ground state jumps from the achiral state $|+\rangle$ to one of the
degenerated chiral states $|\lambda_1\rangle$ and $|\lambda_2\rangle$ at the critical point.
We find that there are two additional higher-energy chiral solutions $|\lambda_{3}\rangle$ and $|\lambda_{4}\rangle$ with $|\lambda_{3}\rangle=\hat{\mathcal{T}}|\lambda_{4}\rangle$ in the
region $f^{\prime}_{cr}<f<f_{cr}$ in the case of $\mathcal{G}=0.4$.

\begin{figure}[h]
  \centering
  \includegraphics[width=0.7\columnwidth]{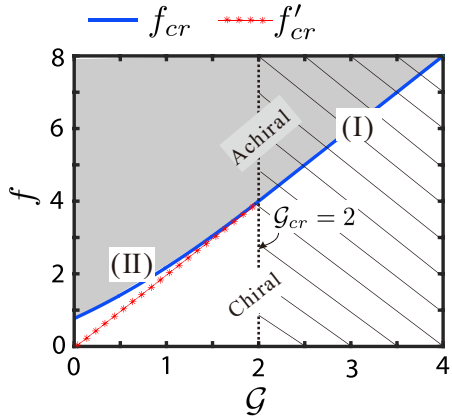}\\
  \caption{Phase diagram of the achiral-chiral transitions in the $(\mathcal{G}$--$f)$ plane. The line of $f_{cr}$ divides the plane
  into upper and lower halves, where the mean-field ground states
  are the achiral and chiral states, respectively. The intersection of the lines of $f_{cr}$ and $f^{\prime}_{cr}$ gives $\mathcal{G}_{cr}=2$. It divides the achiral-chiral transitions into categories (I) and (II). In category (I) where $\mathcal{G}\geq \mathcal{G}_{cr}$, the achiral-chiral transitions present
  typical behaviours as shown in Fig.~\ref{Fig3}~(a) and Fig.~\ref{Fig3}~(b).
  In category (II) where $\mathcal{G}< \mathcal{G}_{cr}$, the achiral-chiral transitions present typical behaviours as shown in Fig.~\ref{Fig3}~(c) and Fig.~\ref{Fig3}~(d).
  }\label{Fig2}
\end{figure}

In Fig.~\ref{Fig2}, {we give phase diagram in
the $(\mathcal{G}$--$f)$ plane to show how $f_{cr}$ and $f^{\prime}_{cr}$
vary with $\mathcal{G}$.}
The line of $f_{cr}(\mathcal{G})$
divides the plane into the upper and lower halves.
The achiral states $|\pm\rangle$ are always the mean-field eigenstates of the system with eigenvalues $\eta_{\pm}=\mp\omega/2$. In the lower half plane, the
system has two degenerated chiral eigenstates $|\lambda_1\rangle$ and $|\lambda_2\rangle$ with lower eigenvalues than $|\pm\rangle$.
In the area surrounded by $f^{\prime}_{cr}(\mathcal{G})$, $f_{cr}(\mathcal{G})$ and $f$-axis, there are further two degenerated chiral eigenstates $|\lambda_{3}\rangle$ and $|\lambda_{4}\rangle$ whose eigenvalues are lower than that of $|\pm\rangle$ but higher than that of $|\lambda_1\rangle$ and $|\lambda_2\rangle$.
The intersection of the lines of $f_{cr}(\mathcal{G})$ and $f^{\prime}_{cr}(\mathcal{G})$ give the critical value $\mathcal{G}_{cr}=2$ labeled with the vertical black dashed line. It divides the achiral-chiral
transitions into categories (I) and (II).
In category (I) where $\mathcal{G}\ge\mathcal{G}_{cr}$, the mean-field ground state changes continuously from the achiral state $|+\rangle$ to one of the two degenerated chiral states $|\lambda_{1}\rangle$ and $|\lambda_{2}\rangle$
with the decrease of $f$. In category (II) where $\mathcal{G}<\mathcal{G}_{cr}$, the mean-field ground state changes discontinuously at the critical point.

{{As we have pointed out, $\mathcal{G}$ is }determined by the species of  {polar} chiral molecules. Different species of  {polar} chiral molecules of $\mathcal{C}_{1}$ symmetry can be classified with the help of Fig.~\ref{Fig2} in two categories. Here, we take propylene oxide and solketal as examples. They are {polar} chiral molecules of $\mathcal{C}_1$ symmetry.
Propylene oxide with $\mathcal{G}\simeq2.7$ can be classified into category (I) { where the} mean-field ground state changes continuously at the critical point. Solketal with $\mathcal{G}\simeq0.46$ can be classified into category (II) {where the} mean-field
ground state changes discontinuously at the critical point.}

\section{Summary}

Starting from the many-body Hamiltonian of the gases of  {polar chiral molecules} with electric dipole-dipole interactions of standard form, we have given the static {non-linear Schr\"{o}dinger} equations in the space of the  {chirality degree of freedom} to explore the achiral-chiral transitions. {Our approach can be applied to {polar} chiral molecules of both $\mathcal{C}_1$ and $\mathcal{C}_2$ symmetry without free parameters.  For the achiral-chiral transitions, we have given the mean-field phase diagram {in the ($\mathcal{G}$--$f$) plane}. We find that, for  {polar} chiral molecules with $\mathcal{G}=(\mu^{m}_{\perp}/\mu^{m}_{z})^2>2$ the mean-field ground state changes continuously from the achiral state $|+\rangle$ to one of the two degenerated chiral state $|\lambda_{1}\rangle$ and $|\lambda_{2}\rangle$ with the decrease of $f$, for {polar} chiral molecules with $\mathcal{G}<2$ the mean-field
ground state changes discontinuously.
This is different from the results predicted with the models
in Refs.~\cite{JCP.112.8743,PRA.91.022709,PRL.88.123001}, where the mean-field ground state always changes continuously.} 
 {It is worthy to note that our discussions based on the molecular electric dipole-dipole interactions are not available to the non-polar chiral molecules, i.e. molecules that belong to the $D_{n}$ point groups~\cite{EN1,EN2}. }

\section{Acknowledgement}
This work was supported by the National Key R\&D
Program of China grant 2016YFA0301200, the National Natural Science Foundation of China
(under Grants No.~11774024, No.~11534002, No.~U1530401, and No.~U1730449), and
the Science Challenge Project (under Grant No. TZ2018003).\\

{}

\end{document}